# Penalized likelihood inference for beta-binomial meta-analysis of proportions of rare events

**Kotaro Sasaki[a,*], Hisashi Noma[a,b]**

[a] The Graduate Institute for Advanced Studies, The Graduate University for Advanced Studies, Tokyo, Japan

[b] Department of Interdisciplinary Statistical Mathematics, The Institute of Statistical Mathematics, Tokyo, Japan

*Corresponding author: Kotaro Sasaki

The Graduate Institute for Advanced Studies

The Graduate University for Advanced Studies

10-3 Midori-cho, Tachikawa, Tokyo 190-8562, Japan

TEL: +81-50-5533-8514

e-mail: sasaki.kotaro.ac@gmail.com

**Abstract**

Meta-analyses of proportions often involve sparse event counts and zero-event studies. The beta-binomial model has been used as a flexible random-effects model for pooling overdispersed and rare-event proportions. However, the ordinary maximum likelihood estimator (MLE) may suffer from finite-sample bias when few studies are available or the mean event probability is close to the boundary. In this article, we propose a maximum penalized likelihood estimator based on the Jeffreys-prior penalty, which has shown favorable finite-sample bias and stability properties in sparse-data models. We also develop Wald-type and profile penalized likelihood confidence intervals (CIs). In a simulation study, the proposed estimator generally achieved higher convergence rates, lower bias, and lower root mean squared error than the ordinary MLE, particularly with fewer studies or lower event probabilities. Additionally, the profile penalized likelihood CIs maintained coverage close to the nominal level. The proposed method provides a useful alternative for meta-analyses of rare-event proportions.

## 1. Introduction

Meta-analyses of proportions are widely used to synthesize event probabilities, such as disease prevalence, adverse event rates, and treatment response rates, across independent studies (Barker et al., 2021; Nyaga et al., 2014). In many applications, the event of interest is rare, resulting in sparse event counts and frequent zero-event studies (Ma et al., 2016; Xu et al., 2021). Conventional random-effects models based on a normal approximation to logit-transformed proportions may perform poorly in such settings. When event counts are small, the normal approximation for the within-study distribution may be inaccurate; moreover, zero-event studies require exclusion or continuity corrections, potentially distorting pooled estimates and confidence intervals (Lin and Xu, 2020; Ma et al., 2016). The *Cochrane Handbook* similarly cautions that methods relying on large-sample approximations may be unsuitable for rare-event meta-analysis (Deeks et al., 2024).

Likelihood-based models provide a natural alternative by directly modeling the binomial counts (Nyaga et al., 2014). Among these, the beta-binomial model has been used as a flexible random-effects model for synthesizing overdispersed and rare-event binary data, while accommodating zero-event studies without ad hoc continuity corrections (Kuss, 2015; Ma et al., 2016; Young-Xu and Chan, 2008). In frequentist applications, the model parameters are commonly estimated by maximum likelihood. However, maximum likelihood estimators (MLEs) are generally subject to finite-sample bias (Firth, 1993). In beta-binomial meta-analysis, such bias may be particularly relevant when the number of studies is small or the mean event probability is close to the boundary, both of which are common in meta-analyses of rare-event proportions.

In this article, we propose a maximum penalized likelihood estimator (MPLE) for the beta-binomial random-effects model in meta-analysis of proportions. The proposed estimator is obtained by maximizing the log-likelihood augmented by the Jeffreys-prior penalty. In canonically parameterized exponential-family models, this MPLE coincides with Firth's mean bias-reduced estimator, which removes the first-order term in the asymptotic bias of the MLE (Firth, 1993). Although this equivalence does not hold for the beta-binomial model, for which mean bias-reduced estimators based on adjusted score equations have been developed (Kenne Pagui et al., 2020), the MPLE has demonstrated favorable finite-sample bias and stability properties across a range of noncanonical sparse-data models (Alam et al., 2022; Kosmidis and Firth, 2021; Xu and Bull, 2023). In addition, the penalized likelihood formulation allows the same objective function to be used for both point estimation and profile inference. Accordingly, we provide both Wald-type confidence intervals (CIs) and profile penalized likelihood CIs. The latter may provide more reliable inference, particularly when the number of studies or the total number of events is small (Heinze, 2006).

Although related penalized likelihood approaches have also been successfully applied in other meta-analytic settings (Evrenoglou et al., 2022; Kosmidis et al., 2017), to our knowledge, this framework has not previously been investigated for beta-binomial meta-analysis of proportions. To assess the performance of the proposed method in this setting, we conduct a simulation study. The proposed MPLE is compared with the ordinary MLE across scenarios involving small numbers of studies, very low event probabilities, and frequent zero-event studies.

## 2. Methods

### 2.1. Beta-binomial model for meta-analysis of proportions

We first describe the beta-binomial model for meta-analysis of proportions. In meta-analysis of proportions, we estimate a pooled event probability from $N$ independent studies. In study $i$ $(i = 1, 2, \dots, N)$, let $Y_i$ denote the number of events and $n_i$ the corresponding sample size. The beta-binomial model is defined by

$$Y_i \mid p_i \sim \text{Binomial}(n_i, p_i),$$

$$p_i \sim \text{Beta}(\alpha, \beta), \tag{1}$$

where $\alpha > 0$, $\beta > 0$, and $p_i$ denotes the event probability for the $i$th study. The probability mass function of $Y_i$ is

$$\Pr(Y_i = y_i) = \binom{n_i}{y_i} \frac{B(y_i + \alpha, n_i - y_i + \beta)}{B(\alpha, \beta)},$$

where $B(\cdot,\cdot)$ denotes the beta function. Let $\pi_i(y_i)$ denote the probability mass function defined above.

We use the following mean-precision parameterization: $\mu = \alpha/(\alpha + \beta)$ and $\kappa = \alpha + \beta$, so that $\alpha = \mu\kappa$ and $\beta = (1 - \mu)\kappa$. Here, $\mu$ is the pooled mean event probability and $\kappa$ is a precision parameter. The corresponding overdispersion parameter is $\rho = 1/(\kappa + 1)$. Under this parameterization, the expectation and variance of $Y_i$ can be expressed as $\mathrm{E}[Y_i] = n_i\mu$ and $\mathrm{Var}[Y_i] = n_i\mu(1 - \mu)\{1 + (n_i - 1)\rho\}$. Thus, $\rho > 0$ represents extra-binomial variation across studies, and as $\rho$ approaches zero, equivalently as $\kappa$ approaches infinity, the beta-binomial model approaches a fixed-effect binomial model.

We define the proposed estimator under the following parameterization:

$$\eta = \log\{\mu/(1 - \mu)\},$$

$$\zeta = \log(\kappa),$$

and let $\boldsymbol{\theta} = (\eta, \zeta)^T \in \mathbb{R}^2$, which allows unconstrained optimization. The log-likelihood for $\boldsymbol{\theta}$ is

$$l(\boldsymbol{\theta}) = \sum_{i=1}^{N} l_i(\boldsymbol{\theta}),$$

where

$$l_i(\boldsymbol{\theta}) = \log \binom{n_i}{y_i} + \log \Gamma\,(y_i + \alpha) + \log \Gamma\,(n_i - y_i + \beta) + \log \Gamma\,(\kappa)$$
$$- \log \Gamma\,(n_i + \kappa) - \log \Gamma\,(\alpha) - \log \Gamma\,(\beta)$$

and $\Gamma(\cdot)$ denotes the gamma function. Then, the ordinary MLE is defined as

$$\widehat{\boldsymbol{\theta}}_{\mathrm{ML}} = \arg\max_{\boldsymbol{\theta}} l(\boldsymbol{\theta}).$$

The score function is

$$U(\boldsymbol{\theta}) = \frac{\partial l(\boldsymbol{\theta})}{\partial \boldsymbol{\theta}} = \sum_{i=1}^{N} u_i(\boldsymbol{\theta}).$$

Here, $u_i(\boldsymbol{\theta}) = \left(u_{\eta i}(\boldsymbol{\theta}), u_{\zeta i}(\boldsymbol{\theta})\right)^T$ is the contribution of study $i$ to the score vector, and

$$u_{\eta i}(\boldsymbol{\theta}) = \frac{\partial l_i(\boldsymbol{\theta})}{\partial \eta} = \kappa\mu(1-\mu)\left(d_{\alpha i}(\boldsymbol{\theta}) - d_{\beta i}(\boldsymbol{\theta})\right),$$

$$u_{\zeta i}(\boldsymbol{\theta}) = \frac{\partial l_i(\boldsymbol{\theta})}{\partial \zeta} = \alpha d_{\alpha i}(\boldsymbol{\theta}) + \beta d_{\beta i}(\boldsymbol{\theta}),$$

where

$$d_{\alpha i}(\boldsymbol{\theta}) = \psi(y_i + \alpha) + \psi(\kappa) - \psi(n_i + \kappa) - \psi(\alpha),$$

$$d_{\beta i}(\boldsymbol{\theta}) = \psi(n_i - y_i + \beta) + \psi(\kappa) - \psi(n_i + \kappa) - \psi(\beta),$$

and $\psi(\cdot)$ denotes the digamma function.

The Fisher information matrix is expressed as $I(\boldsymbol{\theta}) = \sum_{i=1}^{N} \mathrm{E}[u_i(\boldsymbol{\theta})u_i(\boldsymbol{\theta})^T]$. Let $I_i(\boldsymbol{\theta}) = \mathrm{E}[u_i(\boldsymbol{\theta})u_i(\boldsymbol{\theta})^T]$ be the contribution of study $i$ to the information matrix. Because $Y_i$ has finite support $\{0, 1, \dots, n_i\}$, the Fisher information matrix can be computed exactly as

$$I(\boldsymbol{\theta}) = \sum_{i=1}^{N} I_i(\boldsymbol{\theta}) = \sum_{i=1}^{N} \sum_{y=0}^{n_i} \pi_i(y; \boldsymbol{\theta}) u_i(\boldsymbol{\theta}) u_i(\boldsymbol{\theta})^T , \tag{2}$$

where $\pi_i(y; \boldsymbol{\theta}) = Pr(Y_i = y)$.

### 2.2. Maximum penalized likelihood estimation and inference

We consider maximum penalized likelihood estimation based on the Jeffreys-prior penalty. Using the Fisher information matrix in Eq. (2), the penalized log-likelihood is defined as

$$l_p(\boldsymbol{\theta}) = l(\boldsymbol{\theta}) + \frac{1}{2} \log |I(\boldsymbol{\theta})|.$$

The proposed MPLE is then defined by

$$\widehat{\boldsymbol{\theta}}_{\mathrm{MPL}} = \underset{\boldsymbol{\theta}}{\operatorname{argmax}} \, l_p(\boldsymbol{\theta}) \, . \tag{3}$$

Equivalently, $\widehat{\boldsymbol{\theta}}_{\mathrm{MPL}}$ can be obtained as the solution of the adjusted score equations

$$U_p(\boldsymbol{\theta}) = \begin{pmatrix} U_\eta(\boldsymbol{\theta}) + A_\eta(\boldsymbol{\theta}) \\ U_\zeta(\boldsymbol{\theta}) + A_\zeta(\boldsymbol{\theta}) \end{pmatrix} = \mathbf{0},$$

where

$$A_r(\boldsymbol{\theta}) = \frac{1}{2} \mathrm{tr} \left( I(\boldsymbol{\theta})^{-1} \frac{\partial I(\boldsymbol{\theta})}{\partial \boldsymbol{\theta}_r} \right), r \in \{\eta, \zeta\}.$$

In the proposed implementation, we directly maximize the penalized log-likelihood $l_p(\boldsymbol{\theta})$ in Eq. (3). The maximization can be performed using a standard optimization algorithm, such as the BFGS method. The penalized likelihood estimates of the pooled event probability $\hat{\mu}_{\mathrm{MPL}}$ and the precision parameter $\hat{\kappa}_{\mathrm{MPL}}$ are obtained by inverse transformation.

Under standard regularity conditions and for fixed finite true parameter values, the MPLE has the same first-order asymptotic distribution as the ordinary MLE. Therefore, we estimate its covariance matrix using the expected Fisher information of the unpenalized beta-binomial log-likelihood in Eq. (2), evaluated at the MPLE:

$$\widehat{\mathrm{Var}}\left(\widehat{\boldsymbol{\theta}}_{\mathrm{MPL}}\right) = I\left(\widehat{\boldsymbol{\theta}}_{\mathrm{MPL}}\right)^{-1}.$$

The standard error of $\hat{\mu}_{\mathrm{MPL}}$ can be obtained by the delta method, $\widehat{\mathrm{SE}}(\hat{\mu}_{\mathrm{MPL}}) = \hat{\mu}_{\mathrm{MPL}}(1-\hat{\mu}_{\mathrm{MPL}})\sqrt{\left[\widehat{\mathrm{Var}}\left(\widehat{\boldsymbol{\theta}}_{\mathrm{MPL}}\right)\right]_{\eta\eta}}$. Additionally, the lower and upper limits of Wald-type $100(1-\alpha)\%$ CI for $\mu$ are given by $\mathrm{expit}\left(\hat{\eta}_{\mathrm{MPL}} \pm z_{1-\alpha/2}\sqrt{\left[\widehat{\mathrm{Var}}\left(\widehat{\boldsymbol{\theta}}_{\mathrm{MPL}}\right)\right]_{\eta\eta}}\right)$, where $z_{1-\alpha/2}$ denotes the $1-\alpha/2$ quantile of the standard normal distribution.

Alternatively, we construct a profile penalized likelihood CI for $\eta$, which may provide more reliable inference in sparse data settings (Heinze, 2006). The profile penalized log-likelihood is defined as

$$\tilde{l}_P(\eta) = \sup_{\zeta} l_P(\eta, \zeta),$$

where $\zeta$ is treated as a nuisance parameter. The penalized likelihood-ratio statistic is

$$LR_P(\eta) = 2\left\{l_P\left(\hat{\eta}_{\mathrm{MPL}}, \hat{\zeta}_{\mathrm{MPL}}\right) - \tilde{l}_P(\eta)\right\}.$$

Using the first-order chi-squared approximation, a $100(1-\alpha)\%$ CI for $\eta$ is defined as

$$\{\eta : LR_P(\eta) \leq \chi^2_{1,1-\alpha}\},$$

where $\chi^2_{1,1-\alpha}$ is the $1-\alpha$ quantile of the chi-squared distribution with one degree of freedom. The corresponding CI for $\mu$ is obtained by applying the expit transformation to the lower and upper limits.

## 3. Simulation

### 3.1. Simulation settings

To evaluate the performance of the proposed method in comparison with the ordinary MLE,

we conducted a simulation study. Simulation data were generated from a beta-binomial model in Eq. (1) to represent meta-analyses of proportions. For each study, the true event probability $p_i$ was generated from a beta distribution, and the number of events $y_i$ was then generated conditional on $p_i$. The study-specific sample size $n_i$ was generated from a discrete uniform distribution ranging from 100 to 500. The mean of the beta distribution was set to $\mu = 0.005, 0.01,$ or $0.05$, and the corresponding values of $\alpha$ and $\beta$ were used as the beta distribution parameters under $\rho = 0.01$. The number of studies was set to $N = 5, 10, 15,$ or $20$. Thus, a total of 12 scenarios were considered, and 2000 datasets were generated for each scenario. Because datasets with no observed events across all studies would rarely be analyzed in practice, any simulated dataset with zero total events was discarded and resampled until at least one event was observed.

We compared the proposed MPLE based on the beta-binomial model with the MLE based on the same model. For the MLE, 95% CIs were constructed using the Wald and ordinary profile likelihood methods; for the MPLE, they were constructed using the Wald and profile penalized likelihood methods. The performance measures were convergence rate, bias, and root mean squared error (RMSE) for the logit-transformed pooled proportion $\eta$, as well as the coverage probability of the 95% CIs. For each method, bias, RMSE, and coverage were calculated using replications in which that method converged.

### 3.2. Simulation results

We present the simulation results in **Table 1**. Overall, the proposed MPLE performed better than MLE, and its advantage was most evident in settings with fewer studies or lower event probabilities. The MPLE achieved comparable or higher convergence rates than the MLE in all scenarios. The proposed estimator also yielded smaller absolute bias and RMSE across all scenarios. For the 95% CIs, the MPLE Wald intervals showed coverage closer to the nominal level than the MLE Wald intervals, while the MPLE profile penalized likelihood intervals maintained coverage close to the nominal level across scenarios.

**Table 1. Simulation results.**

| Scenario | $N$ | $\mu$ | Convergence rate (%) | | Bias[a] | | RMSE[a] | | Coverage probability (%) | | | |
|---|---|---|---|---|---|---|---|---|---|---|---|---|
| | | | MLE | MPLE | MLE | MPLE | MLE | MPLE | MLE Wald | MPLE Wald | MLE PL | MPLE PPL |
| 1 | 5 | 0.005 | 97.6 | 99.5 | −0.226 | 0.109 | 0.811 | 0.590 | 91.6 | 94.0 | 96.2 | 94.5 |
| 2 | | 0.01 | 98.0 | 99.6 | −0.135 | 0.025 | 0.606 | 0.493 | 88.2 | 93.5 | 93.7 | 95.5 |
| 3 | | 0.05 | 96.6 | 99.2 | −0.029 | −0.007 | 0.254 | 0.246 | 87.3 | 89.9 | 91.8 | 94.0 |
| 4 | 10 | 0.005 | 99.4 | 99.6 | −0.172 | −0.001 | 0.625 | 0.505 | 90.5 | 95.4 | 94.9 | 96.9 |
| 5 | | 0.01 | 99.4 | 99.8 | −0.076 | −0.005 | 0.404 | 0.369 | 91.6 | 94.1 | 94.8 | 95.7 |
| 6 | | 0.05 | 98.8 | 99.5 | −0.011 | −0.001 | 0.169 | 0.167 | 91.4 | 92.5 | 93.6 | 94.3 |
| 7 | 15 | 0.005 | 99.7 | 99.9 | −0.082 | 0.020 | 0.470 | 0.422 | 91.8 | 93.7 | 94.2 | 95.2 |
| 8 | | 0.01 | 99.8 | 99.8 | −0.045 | 0.002 | 0.317 | 0.302 | 92.9 | 93.3 | 94.6 | 94.4 |
| 9 | | 0.05 | 99.7 | 99.7 | −0.012 | −0.005 | 0.138 | 0.136 | 92.9 | 93.4 | 94.8 | 94.6 |
| 10 | 20 | 0.005 | 99.8 | 99.9 | −0.056 | 0.018 | 0.391 | 0.364 | 92.5 | 94.1 | 94.4 | 95.4 |
| 11 | | 0.01 | 99.9 | 99.9 | −0.040 | −0.005 | 0.278 | 0.266 | 93.0 | 93.7 | 94.3 | 94.9 |
| 12 | | 0.05 | 99.6 | 99.8 | −0.004 | 0.001 | 0.121 | 0.120 | 92.5 | 92.6 | 93.6 | 93.7 |

Abbreviations: MLE, maximum likelihood estimator; MPLE, maximum penalized likelihood estimator; Wald, Wald-type confidence interval; PL, profile likelihood confidence interval; PPL, profile penalized likelihood confidence interval.

[a] Bias and RMSE were calculated for the logit-transformed pooled proportion $\eta$; $\mu = 0.005$, $0.01$, and $0.05$ correspond to $\eta = -5.293$, $-4.595$, and $-2.944$, respectively.

## 4. Conclusion

The beta-binomial model provides an effective framework for meta-analyses of proportions, particularly for synthesizing overdispersed and rare-event proportions. However, finite-sample bias of the MLE can be a practical concern when the number of studies is small or the mean event probability is close to the boundary. We proposed Jeffreys-prior penalized likelihood estimation for the beta-binomial random-effects model, together with corresponding Wald-type and profile penalized likelihood CIs. In the simulation study, the proposed MPLE generally showed lower bias and RMSE than the ordinary MLE. Additionally, the profile penalized likelihood CIs maintained coverage close to the nominal level. The proposed method may provide a useful alternative to the ordinary MLE for meta-analyses of proportions.

## Acknowledgements

KS and HN were funded by Grants-in-Aid for Scientific Research from the Japan Society for the Promotion of Science (grant numbers: JP23K11931, JP22H03554, and JP24K21306).

## Declaration of generative AI in the manuscript preparation process

During the preparation of this work the authors used ChatGPT to assist with the English-language editing and code development. After using this tool, the authors reviewed and edited the content as needed and take full responsibility for the content of the published article.

## Data availability

The R code used to implement the proposed method and reproduce the simulation study is available in the GitHub repository at https://github.com/KSasaki7/bbMetaProp.